\documentstyle[12pt]{article}

\input amssym.tex

\begin{document}

\newcommand{\cA}{{\cal A}}
\newcommand{\cB}{{\cal B}}
\newcommand{\cC}{{\cal C}}
\newcommand{\cD}{{\cal D}}
\newcommand{\cE}{{\cal E}}
\newcommand{\cF}{{\cal F}}
\newcommand{\cG}{{\cal G}}
\newcommand{\cH}{{\cal H}}
\newcommand{\cI}{{\cal I}}
\newcommand{\cJ}{{\cal J}}
\newcommand{\cK}{{\cal K}}
\newcommand{\cL}{{\cal L}}
\newcommand{\cM}{{\cal M}}
\newcommand{\cN}{{\cal N}}
\newcommand{\cO}{{\cal O}}
\newcommand{\cP}{{\cal P}}
\newcommand{\cQ}{{\cal Q}}
\newcommand{\cS}{{\cal S}}
\newcommand{\cR}{{\cal R}}
\newcommand{\cT}{{\cal T}}
\newcommand{\cU}{{\cal U}}
\newcommand{\cV}{{\cal V}}
\newcommand{\cW}{{\cal W}}
\newcommand{\cX}{{\cal X}}
\newcommand{\cY}{{\cal Y}}
\newcommand{\cZ}{{\cal Z}}
\def\<{\langle}
\def\>{\rangle}
\def\lgl{\langle\langle}
\def\rgr{\rangle\rangle}
\newcommand{\bra}[1]{\left\langle {#1}\right|}
\newcommand{\ket}[1]{\left| {#1}\right\rangle}
\newcommand{\vev}[1]{\left\langle {#1}\right\rangle}
\newcommand{\be}{\begin{equation}}
\newcommand{\ee}{\end{equation}}
\newcommand{\bn}{\begin{eqnarray}}
\newcommand{\en}{\end{eqnarray}}
\newcommand{\bnn}{\begin{eqnarray*}}
\newcommand{\enn}{\end{eqnarray*}}
\newcommand{\e}{\label}
\newcommand{\nbr}{\nonumber\\[2mm]}
\newcommand{\r}[1]{(\ref{#1})}
\newcommand{\refp}[1]{\ref{#1}, page~\pageref{#1}}
\renewcommand {\theequation}{\thesection.\arabic{equation}}
\renewcommand {\thefootnote}{\fnsymbol{footnote}}
\newcommand{\qq}{\qquad}
\newcommand{\qqq}{\quad\quad}
\newcommand{\biz}{\begin{itemize}}
\newcommand{\eiz}{\end{itemize}}
\newcommand{\ben}{\begin{enumerate}}
\newcommand{\een}{\end{enumerate}}
\def\nc{noncommutative }
\def\ncy{noncommutativity }
\def\com{commutative }
\def\P{Poincar\'e }
\def\th{$\theta_{\mu\nu}$}
\def \simlt{\stackrel{<}{{}_\sim}}
\def\ss{$\theta_{0i}=0$}
\thispagestyle{empty}
\begin{flushright}
\end{flushright}

\begin{center}

{\Large{\bf{On a Lorentz-Invariant Interpretation\\
of Noncommutative Space-Time\\
\vskip .3 cm and Its Implications on Noncommutative QFT}}} \vskip
.7cm

{\bf{\large{M. Chaichian$^a$, P. P. Kulish$^b$, K.
Nishijima$^{a,c}$ and A. Tureanu$^a$}}

{\it $^a$High Energy Physics Division, Department of Physical
Sciences,
University of Helsinki\\
\ \ {and}\\
\ \ Helsinki Institute of Physics, P.O. Box 64, FIN-00014
Helsinki, Finland\\
$^b$ St. Petersburg Department of Steklov Mathematical Institute,
Fontanka 27, St. Petersburg 191023, Russia\\
$^c$ Nishina Memorial Foundation, 2-28-45 Honkomagome, Bunkyo-ku,
Tokyo 113-8941, Japan}}

\end{center}

\setcounter{footnote}{0}

{\bf Abstract} By invoking the concept of twisted Poincar\' e
symmetry of the algebra of functions on a Minkowski space-time, we
demonstrate that the noncommutative space-time with the
commutation relations $[x_\mu,x_\nu]=i\theta_{\mu\nu}$, where
$\theta_{\mu\nu}$ is a {\it constant} real antisymmetric matrix,
can be interpreted in a Lorentz-invariant way. The implications of
the twisted Poincar\'e  symmetry on  QFT on such a space-time is
briefly discussed. The presence of the twisted symmetry gives
justification to all the previous treatments within NC QFT using
Lorentz invariant quantities and the representations of the usual
Poincar\'e symmetry.


\newpage
\section{Introduction}

Quantum field theories on noncommutative space-time have been
lately thoroughly investigated, especially after it has been shown
\cite{SW} that they can be obtained as low-energy limits of open
string theory in an antisymmetric constant background field (for
reviews, see \cite{DN}, \cite{Szabo}). However, the issue of the
lack of Lorentz symmetry has remained a challenge to this moment,
since the field theories defined on a space-time with the
commutation relation of the coordinate operators
\be\label{cr}[\hat x_\mu,\hat x_\nu]=i\theta_{\mu\nu}\ ,\ee
where $\theta_{\mu\nu}$ is a {\it constant} antisymmetric matrix,
are obviously not Lorentz-invariant.

In spite of this well-recognized problem, all fundamental issues,
like the unitarity \cite{unit}, causality \cite{causal1}, UV/IR
divergences \cite{uv/ir}, have been discussed in a formally
Lorentz invariant approach, using the representations of the usual
\P algebra. These results have been achieved using the Weyl-Moyal
correspondence, which assigns to every field operator $\phi(\hat
x)$ its Weyl symbol $\phi(x)$ defined on the commutative
counterpart of the \nc space-time. At the same time, this
correspondence requires that products of operators are replaced by
Moyal $\star$-products of their Weyl symbols:
\be\label{corres} \phi(\hat x)\psi(\hat x)\rightarrow
\phi(x)\star\psi(x), \ee
where the Moyal $\star$-product is defined as
\be\label{star}
\phi(x)\star\psi(x)=\phi(x)e^{\frac{i}{2}\theta^{\mu\nu}\frac{\partial}{\partial
x^\mu}\frac{\partial}{\partial y^\nu}}\psi(y)|_{x=y}\ . \ee
Consequently, the commutators of operators are replaced by Moyal
brackets and the equivalent of (\ref{cr}) is
\be\label{cr'}[ x_\mu,x_\nu]_\star\equiv x_\mu\star
x_\nu-x_\nu\star x_\mu=i\theta_{\mu\nu}\ ,\ee

In fact, admitting that noncommutativity should be relevant only
at very short distances, the noncommutativity has been often
treated as a perturbation and only the corrections to first order
in $\theta$ were computed. As a result, the NC QFT was practically
considered Lorentz invariant in zeroth order in $\theta_{\mu\nu}$,
with the first order corrections coming only from the
$\star$-product.

Later the fact that QFT on 4-dimensional NC space-time is
invariant under the $SO(1,1)\times SO(2)$ subgroup of the Lorentz
group was used \cite{LAG} (for several applications, see
\cite{ls}, \cite{vm}, \cite{axiom}, \cite{CT}). However, a serious
problem arises from the fact that the representation content of
the $SO(1,1)\times SO(2)$ subgroup is very different from the
representation content of the Lorentz group: both $SO(1,1)$ and
$SO(2)$ being abelian groups, they have only one-dimensional
unitary irreducible representations and thus no spinor, vector
etc. representations. In this respect, one encounters a
contradiction with previous calculations, in which the
representation content for the NC QFT was assumed to be the one of
the \P group.

In this letter we shall show that indeed the transformation
properties of the NC space-time coordinates $x_\mu$ can still be
regarded as the transformations under the usual \P algebra, with
their representation content identical to the one of the
commutative case. At the same time, the commutation relation
(\ref{cr'}) appears as the consequence of the noncommutativity of
the coproduct (called noncocommutativity) of the twist-deformed
(Hopf) \P algebra when acting on the products of the space-time
coordinates $x_\mu x_\nu$. As a consequence, the QFT constructed
with $\star$-product on such a NC space-time, though it explicitly
violates the Lorentz invariance, possesses the symmetry under the
proper twist-\P algebra.

\setcounter{equation}{0}

\section{Twist deformation of the Poincar\'e algebra}

The usual \P algebra $\cal P$ with the generators $M_{\mu\nu}$ and
$P_\alpha$ has abelian subalgebra of infinitesimal translations.
Using this subalgebra it is easy to construct a twist element of
the quantum group theory \cite{Dr} (for detailed explanations, see
the monographs \cite{ChPr}, \cite{ChD}), which permits to deform
the universal enveloping of the \P algebra $\cal U(\cal
P)$\footnote{For a deformed \P group with twisted classical
algebra, see \cite{CD}.}.

This twist element $\cal F\in \cal U(\cal P)\otimes\cal U(\cal P)$
does not touch the multiplication in $\cal U(\cal P)$, i.e.
preserves the corresponding commutation relations among
$M_{\mu\nu}$ and $P_\alpha$,
\bn\label{p_alg} [P_\mu,P_\nu]&=&0\ ,\cr
[M_{\mu\nu},M_{\alpha\beta}]&=&-i(\eta_{\mu\alpha}M_{\nu\beta}-\eta_{\mu\beta}M_{\nu\alpha}
-\eta_{\nu\alpha}M_{\mu\beta}+\eta_{\nu\beta}M_{\mu\alpha})\ ,\cr
[M_{\mu\nu},P_\alpha]&=&-i(\eta_{\mu\alpha}P_\nu-\eta_{\nu\alpha}P_\mu)\
,\en
with the essential physical implication that the representations
of the algebra $\cal U(\cal P)$ are the same. However, the action
of $\cal U(\cal P)$ in the tensor product of representations is
defined by the coproduct given, {\it in the standard case}, by the
symmetric map (primitive coproduct)
\bn \Delta_0:\cal U(\cal P)\rightarrow \cal U(\cal P)\otimes\cal
U(\cal P)\cr \Delta_0(Y)=Y\otimes 1 +1\otimes Y\ ,
 \en
for all generators $Y \in \cal P$. The twist element $\cal F$
changes the coproduct of $\cal U(\cal P)$ \cite{Dr}
\be \Delta_0(Y)\mapsto\Delta_t(Y)=\cF\Delta_0(Y)\cF^{-1}\ .\ee
This similarity transformation is consistent with all the
properties of $\cal U(\cal P)$ as a Hopf algebra if $\cal F$
satisfies the following twist equation\footnote{See more detailed
explanations in monographs on quantum groups (e.g. \cite{ChPr},
\cite{ChD}).}:
\be\label{te} \cF(\Delta_0\otimes id) \cF=\cF(id\otimes
\Delta_0)\cF\ . \ee
Taking the twist element in the form of an abelian twist \cite{R},
\be \cF = exp(\frac{i}{2} \theta^{\mu\nu}P_\mu\otimes P_\nu)\ ,\ee
one can check that the twist equation (\ref{te}) is valid.

Since the generators of translations $P_\alpha$ are commutative,
their coproduct is not deformed ($\Delta_t=\Delta_0$ is primitive)
\be\label{tp}
\Delta_t(P_\alpha)=\Delta_0(P_\alpha)=P_\alpha\otimes1+1\otimes
P_\alpha\ .\ee
However, the coproduct of the Lorentz algebra generators is
changed:
\be\Delta_t(M_{\mu\nu})=Ad\
e^{\frac{i}{2}\theta^{\alpha\beta}P_\alpha\otimes
P_\beta}\Delta_0(M_{\mu\nu})=e^{\frac{i}{2}\theta^{\alpha\beta}P_\alpha\otimes
P_\beta}\Delta_0(M_{\mu\nu})e^{-\frac{i}{2}\theta^{\alpha\beta}P_\alpha\otimes
P_\beta}\ .\ee
Using the operator formula $Ad\ e^BC=e^B\ C\
e^{-B}=\sum\limits_{n=0}^\infty\frac{1}{n!}\underbrace{[B,[B,...[}_n
B,C]]=\sum\limits_{n=0}^\infty\frac{(Ad B)^n}{n!}\ C$ and the
commutation relation between $M_{\mu\nu}$ and $P_\alpha$ (last
line of (\ref{p_alg})), we obtain the explicit form of the
coproduct\footnote{After the submission of the present work to the
hep-th Archive, we were informed that the result (\ref{tm})
appears also in \cite{Wess}, which is an extended version of the
talk given by Julius Wess in the "Balkan Workshop 2003".}
$\Delta_t(M_{\mu\nu})$:
\bn\label{tm} \Delta_t(M_{\mu\nu})&=&Ad\
e^{\frac{i}{2}\theta^{\alpha\beta}P_\alpha\otimes
P_\beta}\Delta_0(M_{\mu\nu})\cr &=&M_{\mu\nu}\otimes 1+1 \otimes
M_{\mu\nu}-\frac{1}{2}\theta^{\alpha\beta}[(\eta_{\alpha\mu}
P_\nu-\eta_{\alpha\nu}P_\mu)\otimes P_\beta\cr
&+&P_\alpha\otimes(\eta_{\beta\mu}P_\nu-\eta_{\beta\nu}P_\mu)]\
.\en

It is known (cf. \cite{ChPr}, \cite{KM}) that having a
representation of a Hopf algebra $\cH$ in an associative algebra
$\cA$ consistent with the coproduct $\Delta$ of $\cH$ (a Leibniz
rule)
\be h(a\cdot b)=h_{1}(a)\cdot h_{2}(b)\ ,\ \ \ \Delta(h)=h_{
1}\otimes h_{2}\ ,\ee
the multiplication in $\cA$ has to be changed after twisting
$\cH$. The new product of $\cA$ consistent with the twisted
coproduct $\Delta_t$ is defined as follows: let $\cF =\sum
f_{1}\otimes f_{2}$, then
\be\label{multipl} a\star b=\sum (\bar f_{1}(a))\cdot(\bar
f_{2}(b))\ ,\ee
where $\bar\cF=\sum\bar f_{1}\otimes\bar f_{2}$ denotes the
representation of $\cF^{-1}$ in $\cA\otimes\cA$,
%
and the action of elements $\bar f\in\cH$ on elements $a,b \in\cA$
is the same as without twisting.

Let us now consider the commutative algebra $\cA$ of functions,
$f(x)$, $g(x)$,..., depending on coordinates $x_\mu$,
$\mu=0,1,2,3$, in the Minkowski space $M$. In $\cA$ we have the
representation of $\cal U(\cal P)$  generated by the standard
representation of the Poincar\' e algebra:
\be\label{pa}  {P}_\mu f(x) =i\partial_\mu f(x)\ ,\ {M}_{\mu\nu}
f(x)=i(x_\mu\partial_\nu - x_\nu\partial_\mu)f(x)\ ,\ee
acting on coordinates as follows:
\be\label{act_coord} {P}_\mu x_\rho=i\eta_{\mu\rho}\ ,\
{M}_{\mu\nu}x_\rho=i(x_\mu\eta_{\nu\rho} -x_\nu\eta_{\mu\rho})\
.\ee
The \P algebra acts on the Minkowski space $x_\mu$, $\mu=0,1,2,3$
with commutative multiplication:
\be m(f(x)\otimes g(x)):=f(x)g(x)\ .\ee

When twisting $\cU(\cP)$, one has to redefine the multiplication
according to (\ref{multipl}), while retaining the action of the
generators of the Poincar\'e algebra on the coordinates as in
(\ref{act_coord}):
\bn\label{star'} m_t(f(x)\otimes g(x))&=:&f(x)\star g(x)=m\circ
e^{-\frac{i}{2}\theta^{\alpha\beta}{P}_\alpha\otimes{P}_\beta}(f(x)\otimes
g(x))\cr
&=&m\circ
e^{\frac{i}{2}\theta^{\alpha\beta}\partial_\alpha\otimes\partial_\beta}(f(x)\otimes
g(x))\ .\en
Specifically, one can now easily compute the commutator of
coordinates:
\bn m_t(x_\mu\otimes x_\nu)&=&x_\mu\star x_\nu=m\circ
e^{-\frac{i}{2}\theta^{\alpha\beta}{P}_\alpha\otimes{P}_\beta}(x_\mu\otimes
x_\nu)\cr&=&m\circ [x_\mu\otimes x_\nu
+\frac{i}{2}\theta^{\alpha\beta}\eta_{\alpha\mu}\otimes\eta_{\beta\nu}
]\cr &=&x_\mu
x_\nu+\frac{i}{2}\theta^{\alpha\beta}\eta_{\alpha\mu}\eta_{\beta\nu}\
,\cr m_t(x_\nu\otimes x_\mu)&=&x_\nu\star x_\mu=x_\nu x_\mu+\frac
{i}{2}\theta^{\alpha\beta}\eta_{\alpha\nu}\eta_{\beta\mu}\ .
 \en
Hence,
\be\label{cr''} [x_\mu,x_\nu]_
\star=\frac{i}{2}\theta^{\alpha\beta}(\eta_{\alpha\mu}\eta_{\beta\nu}-\eta_{\alpha\nu}\eta_{\beta\mu})
=i\theta_{\mu\nu}\ ,\ee
which is indeed the Moyal bracket (\ref{cr'}).

\section{QFT on space-time with twisted Poincar\'e symmetry}

 Comparing (\ref{star}) and (\ref{star'}) (or equivalently
(\ref{cr'}) and (\ref{cr''})), it is obvious that building up the
\nc quantum field theory through Weyl-Moyal correspondence is
equivalent to the procedure of redefining the multiplication of
functions, so that it is consistent with the twisted coproduct of
the Poincar\'e generators (\ref{tp}), (\ref{tm}). The QFT so
obtained is invariant under the twisted Poincar\'e algebra. The
benefit of reconsidering NC QFT in the latter approach is that it
makes transparent the {\it invariance} under the twist-deformed
Poincar\'e algebra, while the first approach highlights the {\it
violation} of the Lorentz group.

To show this invariance, let us take, as an instructive example,
the product $f_{\rho\sigma}(x)=x_\rho x_\sigma$. In the standard
non-twisted case, the action of the Lorentz generators on this
product reads as:
\be\label{li} {M}_{\mu\nu}f_{\rho\sigma}=i(x_\mu\partial_\nu -
x_\nu\partial_\mu)f_{\rho\sigma}
 = i(f_{\mu\sigma}\eta_{\nu\rho}
-f_{\nu\sigma}\eta_{\mu\rho}+f_{\rho\nu}\eta_{\mu\sigma}
-f_{\rho\mu}\eta_{\nu\sigma})\ ,\ee
expressing the fact that $f_{\rho\sigma}$ is a rank-two Lorentz
tensor. In the twisted case, $f_{\rho\sigma}$ should be replaced,
according to (\ref{star'}), by the symmetrized
expression\footnote{We use the symmetrization because, due to the
commutation relation $[x_\mu,x_\nu]_\star=i\theta_{\mu\nu}$ (where
$\theta_{\mu\nu}$ is twisted-\P invariant, as shown also in the
consistency check performed below), every tensorial object of the
form $x_\mu\star x_\nu\star\cdots\star x_\sigma$ can be written as
a sum of symmetric tensors of lower or equal ranks, so that the
basis of the representation algebra ${\cal A}_t$ is symmetric.
This statement is valid in general in the case of the universal
enveloping algebras of Lie algebras.}
$f^t_{\rho\sigma}=x_{\{\rho}\star x_{\sigma\}}=\frac{1}{2}(x_\rho
\star x_\sigma+x_\sigma\star x_\rho)$, and correspondingly the
action of the Lorentz generator should be applied through the
twisted coproduct:
\be {M}^t_{\mu\nu}f^t_{\rho\sigma}={m}_t\circ(\Delta_t
({M}_{\mu\nu})(x_\rho\otimes x_\sigma))\ .\ee
In the above equation, $M^t_{\mu\nu}$ denotes the usual Lorentz
generator, but with the action of a twisted coproduct. A
straightforward calculation gives:
\be\label{li*} {M}^t_{\mu\nu}f^t_{\rho\sigma} =
i(f^t_{\mu\sigma}\eta_{\nu\rho}
-f^t_{\nu\sigma}\eta_{\mu\rho}+f^t_{\rho\nu}\eta_{\mu\sigma}
-f^t_{\rho\mu}\eta_{\nu\sigma})\ ,\ee
which is analogous to (\ref{li}), confirming the (expected)
covariance under the twisted Poincar\'e algebra. This argument
extends to any symmetrized tensor formed from the $\star$-products
of $x$'s. For example, the invariance of Minkowski length
$s^2_t=x_\mu\star x^\mu=x_\mu x^\mu$ is obvious: multiplying
(\ref{li*}) by $\eta^{\rho\sigma}$, one obtains
${M}^t_{\mu\nu}s^2_t=0$.

As a consistency check, we shall calculate the action of
$M^t_{\mu\nu}$ on the antisymmetric combination $2x_{[\rho}\star
x_{\sigma]}=[x_\rho,x_\sigma]_\star$:
\bn
{M}^t_{\mu\nu}([x_\rho,x_\sigma]_\star)&=&([x_\mu,x_\sigma]_\star-i\theta_{\mu\sigma})\eta_{\nu\rho}
-([x_\nu,x_\sigma]_\star-i\theta_{\nu\sigma})\eta_{\mu\rho}\cr
&-&([x_\mu,x_\rho]_\star-i\theta_{\mu\rho})\eta_{\nu\sigma}
+([x_\nu,x_\rho]_\star-i\theta_{\nu\rho})\eta_{\mu\sigma}=0\ . \en
Thus, we have $M^t_{\mu\nu}\theta_{\rho\sigma}=0$, since
$\theta_{\rho\sigma}=-i[x_\rho,x_\sigma]_\star$, i.e. the
antisymmetric tensor $\theta_{\rho\sigma}$ is twisted-\P
invariant.

Therefore, the Lagrangian obtained by replacing all the usual
products of fields in the corresponding commutative theory with
$\star$-products, though it breaks the Lorentz invariance in the
usual sense, it is, however, invariant under the twist-deformed \P
algebra.

Another important feature of the QFT with twist-deformed
Poincar\'e symmetry deserves a special highlighting: the
representation content of the NC QFT is exactly the same as for
its commutative correspondent. It is easy to see that the action
of the Pauli-Ljubanski operator, $W_\alpha=-\frac{1}{2}\epsilon
_{\alpha\beta\gamma\delta}M^{\beta\gamma}P^\delta$ is not changed
by the twist (due to the commutativity of the translation
generators) and $P^2$ and $W^2$ retain their role of Casimir
operators. Consequently, the representations of the twisted \P
algebra will be, just as in the commutative case, classified
according to the eigenvalues of these invariant operators, $m^2$
and $m^2s(s+1)$, respectively. Besides justifying the validity of
the results obtained so far in NC QFT using the representations of
the \P algebra, this aspect will cast a new light on other
closely-related fundamental issues, such as the CPT and the
spin-statistics theorems in NC QFT \cite{vm,axiom,CNT}.

\section{Conclusions}
In this letter we have shown that the quantum field theory on NC
space-time possesses symmetry under a twist-deformed \P algebra.
The twisted Poincar\' e symmetry exists provided that: (i) we
consider $\star$-products among functions instead of the usual one
and (ii) we take the proper action of generators specified by the
twisted coproduct. As a byproduct with major physical
implications, the representation content of NC QFT, invariant
under the twist-deformed \P algebra, is identical to the one of
the corresponding commutative theory with usual \P symmetry. Some
of the applications of the present treatment of the symmetry
properties of NC QFT will be considered in a forthcoming
communication \cite{Haag}.

 \vskip 0.3cm {\Large\bf{Acknowledgements}}

We are indebted to Peter Pre\v{s}najder for many illuminating
discussions, remarks and useful suggestions.

The financial support of the Academy of Finland under the Projects
No. 54023 and 104368 is greatly acknowledged. The work of PPK was
partly supported by the RFBR grant 03-01-00593.

\end{document}